\documentstyle[epsf]{laa}     

\newcommand{\be}{\begin{equation}}
\newcommand{\ee}{\end{equation}}
\newcommand{\etal}{{\it et al.}}
\newcommand{\hmp}{h^{-1}Mpc}
\newcommand{\bef}{\begin{figure}}
\newcommand{\eef}{\end{figure}}

\def\spose#1{\hbox to 0pt{#1\hss}}
\def\ltapprox{\mathrel{\spose{\lower 3pt\hbox{$\mathchar"218$}}
 \raise 2.0pt\hbox{$\mathchar"13C$}}}
\def\gtapprox{\mathrel{\spose{\lower 3pt\hbox{$\mathchar"218$}}
 \raise 2.0pt\hbox{$\mathchar"13E$}}}
\def\inapprox{\mathrel{\spose{\lower 3pt\hbox{$\mathchar"218$}}
 \raise 2.0pt\hbox{$\mathchar"232$}}}

\begin{document}
   \thesaurus{07         
              (07.09.1;  
               03.01.1;  
               04.01.1)  
             }

\title{Scale-invariant properties of the APM-Stromlo  survey}
   \subtitle{}

   \author{F. Sylos Labini \inst{1,2} and  M. Montuori \inst{1}}
   \institute{ 
       	      INFM Sezione Roma1,       
	      Dip. di Fisica, Universit\'a "La Sapienza",
	      P.le A. Moro, 2, 
              I-00185 Roma, Italy
  \and
              D\'ept.~de Physique Th\'eorique, Universit\'e de Gen\`eve, 
24, Quai E. Ansermet, CH-1211 Gen\`eve, Switzerland
}

\date{Received -- -- --; accepted -- -- --}

\maketitle
 
\begin{abstract}
We investigate the statistical properties of 
the APM-Stromlo redshift survey by using 
the concepts and methods of 
modern Statistical Physics. We 
find that galaxy distribution in 
this survey  exhibits scale invariant
properties with fractal dimension $D = 2.1 \pm 0.1$,
up to $\sim 40 \hmp$, i.e. the limit of its statistical validity,
without any tendency towards homogenization.
No intrinsic characteristic scales are definitely found
in this galaxy sample. We present several tests to study
the statistical reliability of the results. 
\end{abstract}


\section{Introduction}

In the debate about the statistical properties of 
 the available three dimensional galaxy 
samples (see Pietronero \etal 1997 and Davis  1997
for the two different points of view on the subject) 
  it is often claimed (see e.g. Peebles, 1980) 
that there exists a well defined {\it correlation length}
$r_0$ which separates 
a correlated regime ($r < r_0$) from an uncorrelated one 
($r > r_0$). Such a characteristic length 
has been identified by 
the  usual  statistical analysis of spatial galaxy distribution 
through the two points correlation  function 
$\xi(r)$ (Peebles, 1980), and it is 
  defined as $\xi(r_0) \equiv 1$. 
  Although the standard value 
 is estimated to be $r_0 \approx 5 \hmp$
(Davis \& Peebles, 1983; Davis, 1997)
different authors have measured a  value  for $r_0$
larger than $10 \hmp$ (e.g. Benoist \etal, 1996; Park \etal, 1994), the reason 
of  such a disagreement being attributed to a "luminosity bias": galaxies 
of different intrinsic luminosities should have different correlation
lengths. While Davis \etal (1988) found a quantitative result for 
the shift of $r_0$ as a function of   absolute magnitude, 
other authors (e.g. Benoist \etal, 1996; Park \etal, 1994) 
neither have confirmed such a behavior nor have given an
alternative quantitative explanation for such an effect.

Some years ago we criticized this approach and 
proposed a new one based on the 
{\it concepts and methods of 
modern Statistical Physics} (Pietronero 1987; Coleman \& Pietronero 1992;
Pietronero \etal, 1997; Sylos Labini \etal, 1997). 
By using this more general framework, we present in this paper
the analysis of the Stromlo-APM
redshift survey (Loveday \etal, 1992; Loveday \etal, 1996).

The main result is that galaxy distribution in this survey does not
exhibit any characteristic length scale and, on the contrary,
 we find that 
it is characterized by having scale invariant properties 
with fractal dimension $D = 2.1 \pm 0.1$ up to the 
statistical limit of this galaxy sample (i.e. $\sim 40 \hmp$).

The very first consequence 
of this result is that 
the usual statistical methods 
(as for example $\xi(r)$), 
based on the assumption of homogeneity, 
are therefore inconsistent for the analysis
of irregular distribution such as the one 
present in this sample: unless a well defined
cut-off towards homogenization has been identified,
all the results based on the $\xi(r)$, such as
the so-called luminosity segregation effect, are 
artifacts. This is the origin of the confusing statements
about "luminosity segregation". In Sylos Labini \& Pietronero (1996)
we have clarified this point by showing that galaxies 
of different morphology indeed have different clustering properties
(for example ellipticals are mainly in the core of
rich galaxy clusters, while spirals are in the field),
but this fact has no relation with the 
increasing of $r_0$  found in galaxy catalogs (Sylos Labini \etal, 1997).
The segregation of galaxies with different morphology is,
 on the contrary, related to the 
multifractal nature of galaxy distribution, i.e. 
to the correlation between galaxy positions and luminosities.

Our results are therefore in contrast with those of 
Loveday \etal 1996, where a well defined 
correlation length $r_0 \sim 5 \hmp$ has been identified. 
We clarify the reason of such a disagreement, and we 
present various tests to asset the robustness of the present 
analysis.


\section{The sample}
   
The Stromlo-APM Redshift Survey (SARS) consists of 1797 galaxies 
with 
$b_J \leq 17.15$ selected randomly at a rate of 1 in 20 from APM scans
(Loveday \etal, 1992; Loveday \etal, 1996). 
The survey covers a solid angle of
 $\Omega = 1.3 sr$
in the south galactic hemisphere, delimited by
 $21^h \ltapprox \alpha \ltapprox 5^h$ and $-72.5^{\circ} \ltapprox \delta 
\ltapprox -17.5^{\circ}$. 
An important selection effects exists: 
galaxies with apparent magnitude brighter than $b_j=14.5$ 
are not included in the sample because of photographic saturation (see 
Fig.\ref{apmfig1}).
\begin{figure}
\vspace{10cm}
\caption{\label{apmfig1}
Absolute magnitude versus distance diagram for the SARS catalog.
There are shown the two limiting function with apparent magnitude
$m^1_{lim}=14.5$ and $m^2_{lim}=17.15$. The survey is complete up to
$\sim 17-th$ magnitude and does not contain galaxies brighter than $\sim 14.5$.}
\eef
Moreover in  a magnitude limited sample 
another  important selection effect   must be considered:
at every distance in the 
apparent magnitude limited survey, there is a definite limit in intrinsic 
luminosity
which is the absolute magnitude of the faintest galaxy
visible that distance. Hence at large distances, intrinsically
faint objects are not observed whereas at smaller distances they are observed.
In order to analyze the statistical properties of galaxy distribution
without introducing any a priori assumption,
a catalog without
this selection effect
must be used. In general, it exists a very well known procedure to
obtain a sample unbiased by this luminosity selection effect:
this is the so-called {\it  "volume limited"} (VL) sample.
A  VL  sample contains every galaxy in the volume
which is more luminous than a certain limit, so that in such a
 sample
there is no incompleteness for any observational
luminosity selection effect (see for example 
Davis \& Peebles, 1983; Coleman \&
Pietronero, 1992).  

In order to construct VL samples we have adopted two
different procedures. The first is the standard one, i.e. we have 
introduced an upper cutoff in the distance and computed the 
corresponding 
cutoff in absolute magnitude. The characteristics
of these samples are reported in Tab.\ref{tableapm1}.
\begin{table}  
\begin{tabular}{lllll} 
\hline
 Sample & $R_{VL} (\hmp)$  & $M_{lim}$  & N        & $\langle \ell \rangle (\hmp)$ \\
\hline
 VL18   &              107 & -18.0      & 325        & 4.2       \\
 VL19   &              170 & -19.0      & 481        & 6.2       \\
 VL20   &              269 & -20.0      & 403        & 11        \\
       &		  & 	       &                  &          \\
\hline
\end{tabular}
\caption{The VL subsamples of the APM catalog \label{tableapm1}:
$R_{VL}$ is the distance corresponding to the absolute magnitude
limit of the sample $M_{lim}$; $N$ is the number of points contained and 
$\langle\ell \rangle $ is the   average distance between neighbor galaxies. 
}
 \end{table}

The second procedure consists by putting  two limits in distances
and compute the corresponding two limits in absolute magnitude
(see Tab.\ref{tableapm2}, 
where WL12 means the VL sample limited by the distance range 
 $100\div 200  \hmp$). 
In such a way we can avoid the selection effects due to the 
fact that in this survey are not included galaxies brighter 
than $14.5$. 
\begin{table}  
\begin{tabular}{lllll}
\hline
 Sample  & $\Delta R_{VL} $  & $\Delta M_{lim}$ & N   & $\langle \ell \rangle  $    \\
 \hline
  WL12    & $100 \div 200$         & $-19.35 \div -20.5$  & 451    & 7.3  \\
  WL51     & $50 \div 100$ & $-17.85 \div -19.0$  & 170    & 4.6  \\
  WL153     & $150 \div 300$ & $-20.2 \div -21.38$  & 230    & 14.5  \\
 \hline
\end{tabular}
\caption{The VL subsamples of the APM catalog 
with a double cut in distance $\Delta R_{VL} (\hmp)$ and absolute magnitude
$\Delta M_{lim}$ \label{tableapm2}}
  \end{table}
  
  The measured velocities of the galaxies have been
expressed in
the preferred frame of the Cosmic Microwave Background Radiation
(CMBR), i.e. the heliocentric velocities of
galaxies have been corrected
for the solar motion with respect to the CMBR, according
with the formula
$\vec{v} =\vec{v}_{m}+316 cos \theta \; \; km s^{-1}
$
where $\vec{v}$ is the corrected velocity,
$\vec{v}_{m}$ is the observed velocity and $\theta$
is the angle between the observed
velocity and the direction of the CMBR dipole anisotropy
($\alpha=169.5^{\circ}$ and $\delta=-7.5^{\circ}$).
From these corrected velocities, we have calculated
the comoving distances
$r(z)$, with for example $q_0=0.5$, by using the Mattig's relation
(see e.g. Park \etal, 1994)
\be
\label{e47}
r(z)=6000\left(1-\frac{1}{\sqrt{(1+z)}}\right) \hmp \; .
\ee

It is important to stress that the analyses presented here 
have been performed in redshift space. We have not applied
any correction to take into account the effect of
peculiar velocity distortions. However 
we expect that these corrections are negligible on scales larger than
$\sim 10 \hmp$, due to the fact that the amplitude of
peculiar motions is not larger than $\sim 1000 km s^{-1}$.


\section{The Average Conditional Density}

We   start by recalling  the 
concept of correlation. If the presence of an object at the point $r_1$ 
influences the probability of finding another object 
at $r_2$, 
these two points are correlated. Therefore there is a correlation
at  $r$ if, on average
\be
\label{e324}
G(r) = \langle n(0)n(r)\rangle   \ne \langle n\rangle  
\ee
where we average on all occupied points chosen as origin.
On the other hand, there is no correlation if
\be
\label{e325}
G(r) \approx \langle n\rangle  ^2 \; .
\ee
The physically meaningful definition of the homogeneity scale
 $\lambda_0$ 
is therefore the length scale which separates correlated regimes from
uncorrelated ones.

In practice, it  is useful 
to normalize the correlation function (CF) to the size  of the
sample   analyzed. Then we use, following Coleman \& Pietronero (1992)
\be
\label{e326}
\Gamma(r) = \frac{<n(r)n(0)>}{<n>} = \frac{G(r)}{<n>}
\ee
where $\:<n>$ is the average density of the sample.  We stress
that this normalization does not introduce any bias even if the average
density is sample-depth dependent,
as in the case of fractal distributions,
because it represents
only an overall normalizing factor. 
In order to compare results from different catalogs
it is however more useful to use $\Gamma(r)$, in which
the size of a catalog only appears via the combination
$N^{-1}\sum_{i=1}^{N}$, so that a larger sample 
volume only enlarges the statistical sample over which averages are taken.
On the contrary $G(r)$   
has an amplitude that is an explicit function of the sample's size
scale.

$\Gamma(r)$ measures the average density within a spherical shell 
of thickness $\Delta r$ at distance $\:\vec{r}$ from an
occupied point at $\vec{r_i}$, 
and it is called the {\it conditional average density}
 (Coleman \& Pietronero, 1992).
Such a function can 
can be estimated by  
\be
\label{e327}
\Gamma(r) = \frac{1}{N(r)} \sum_{i=1}^{N(r)} \frac{1}{4 \pi r^2 \Delta r}
\int_{r}^{r+\Delta r} n(\vec{r}_i+\vec{r'})d\vec{r'} = 
\frac{BD}{4 \pi} r^{D-3}
\ee
where  $D$ is the fractal dimension and the prefactor $\:B$
is instead related to the lower cut-offs, i.e. 
the lower scale at which the self-similarity is broken 
(Coleman \& Pietronero, 1992; Sylos Labini \etal, 1997).
  The quantity $N(r)$ in eq.\ref{e327} is the actual number of points
 over which the average is done. This is in general a function 
 of the radius $r$ of the shell, because we consider 
 only those points for which the spherical shell of radius
 $r$ in entirely contained in the sample. Clearly as $r$ grows
 the number of points $N(r)$ decreases, and for this reason the 
 determination of $\Gamma(r)$ for large $r$ 
 is more noisy (see below).
 
If the distribution is fractal up to a certain distance $\lambda_0$,
and then it becomes homogeneous,  
  $\Gamma(r)$ has a power law decaying with
distance up to $\lambda_0$, and then it flattens 
towards a constant value.
Hence by studying the behavior of $\Gamma(r)$
it is possible to detect the eventual scale-invariant versus 
homogeneous  properties
of the sample. 

In general, it
 is also very useful to use the {\it integrated conditional density}
\be
\label{e328}
\Gamma^*(r) = \frac{3}{4 \pi r^3} \int_{0}^{r} 4 \pi r'^2 \Gamma(r') dr' =
\frac{3B}{4 \pi} r^{D-3} 
\ee
This function  produces an artificial smoothing of
rapidly varying fluctuations, but it correctly
reproduces global properties (Coleman \& Pietronero, 1992).
For a fractal structure, $\Gamma(r)$ has a power law behavior
and the integrated conditional density is 
\be
\label{e329}
\Gamma^*(r)= \frac{3}{D} \Gamma(r).
\ee
For an homogeneous distribution ($D=3$) these two functions
are exactly the same and equal to the average density.

Contrary to the conditional density,
 the information given by the $\xi(r)$ function
 is biased by the 
a priori (untested) assumption of homogeneity.
 Pietronero  and collaborators 
(Pietronero, 1987; Coleman \& Pietronero, 1992; Sylos Labini \etal, 1997)
 have clarified some crucial points of the
standard correlations analysis, and in particular they have discussed the 
physical meaning
of the so-called {\it "correlation length"}
  $\:r_{0}$ found with the standard
approach   and defined by the relation 
$\xi(r_{0})\equiv 1$
where
\be
\label{e331}
\xi(r) = \frac{<n(\vec{r_{0}})n(\vec{r_{0}}+ \vec{r})>}{<n>^{2}}-1
\ee
is the two points correlation function used in the standard analysis.
The basic point in the present discussion,
is that the mean density $ \langle n \rangle $
used in the normalization of $\:\xi(r)$, is not a well defined quantity
in the case
of self-similar distribution and it is a direct function of the sample size.
Hence only in the case that 
homogeneity  has been reached well within the sample
limits the $\:\xi(r)$-analysis is meaningful, otherwise
the a priori assumption of homogeneity is incorrect and 
characteristic lengths, like $\:r_{0}$, became spurious (see e.g. Sylos 
Labini \etal, 1997  for an exhaustive discussion of the matter).

Given a certain spherical sample with solid angle $\Omega$ and depth $R_s$,
it is important to define which is 
 the maximum distance up to which it 
is possible to compute the correlation function ($\Gamma(r)$ or $\xi(r)$).
As discussed in Coleman \& Pietronero 
(1992), we have limited our analysis to an
effective  depth
$R_{eff}$ that is of the order of the radius of the maximum
sphere fully contained in the sample volume.
 The reason why
$\Gamma(r)$ (or $\xi(r)$) cannot
be computed for $r > R_{eff}$
is essentially the following.
 When one evaluates the correlation
function
(or power spectrum - see Sylos Labini \& Amendola 1996) beyond $R_{eff}$,
then one  makes explicit assumptions on what
lies beyond the sample's boundary.  In fact, even in absence of
corrections for selection effects, one
is forced to consider incomplete shells
calculating $\Gamma(r)$ for $r \gtapprox R_{eff}$,
thereby
implicitly assuming that what one  does not find  in the part of the
shell not included in the sample is equal to what is inside.

The {\it maximum depth} of a reliable statistical analysis, 
is limited by the radius of the sample (as previously discussed), 
while the {\it minimum distance}
depends on the number of points contained in the volume 
and on the fractal dimension. 
For a Poisson distribution the mean average distance between nearest neighbors 
is of the order $\langle\ell \rangle \sim (V/N)^{\frac{1}{3}}$.  
It is possible to compute  the average  distance 
between neighbor galaxies $\langle \ell \rangle$, in a fractal
distribution with dimension $D$, and the result is 
\be
\label{e5}
\langle \ell \rangle = \left(\frac{1}{B}\right)^{\frac{1}{D}} \Gamma
\left(1 + \frac{1}{D} \right)
\ee
where $\Gamma$ is the Euler's gamma-function (Sylos Labini \etal,  1997b). 
Clearly this quantity   is related 
to the lower cut-off of the distribution $B$ (eq.\ref{e327}) 
and to the fractal dimension $D$.
If we measure the conditional density at distances 
$ r \ltapprox \langle \ell \rangle$, 
we are affected 
by a {\it finite size effect}. In fact, due the depletion of points at these 
distances 
we underestimate the real conditional density finding an higher value 
for the correlation exponent (and hence a lower value for the fractal 
dimension). 
In the limiting case at the distance $ r \ll \langle \ell \rangle $, 
we can find almost no points and 
the slope is
$\gamma=-3$ ($D=0$). 
In general, when one  measures $\Gamma(r)$ at distances 
which correspond to 
a fraction of $ \langle \ell \rangle$, 
one finds systematically an higher value of the 
conditional density exponent. 
Such a trend  is completely spurious and due to the depletion of
points at such distances. It is worth to notice that this effect
gives rise to a curved behaviour of 
$\Gamma^*(r)$  (eq.\ref{e329}) at small distances, because 
of its integral nature.
 
An important point in the study
of galaxy distribution, is that galaxies are characterized
by having very different intrinsic luminosities.
In order to take into account this effect,  in what follows
we assume that 
\be
\label{e49}
\nu(L,\vec{r}) = \phi(L) \Gamma(\vec{r}) \; , 
\ee
i.e. that the number of galaxies for unit luminosity
and volume $\nu(L,\vec{r})$ can be expressed as
the product of the space density
$ \Gamma(\vec{r})$  (eq.\ref{e327}) and  the luminosity function $ \phi(L)$
($L$ is the intrinsic luminosity).
This is a crude approximation
in view of the multifractal properties of the distribution
(correlation between position and luminosity).
However, for the purpose of the present discussion,
the previous approximation is rather good and the
explicit consideration of the multifractal properties
 have a minor effect on the properties    
discussed (Sylos Labini \& Pietronero, 1996).
                    
In view of eq.\ref{e49}, to each VL sample (limited by the 
absolute magnitude $M_{VL}$)
we can associate the luminosity factor
\be
\label{e410}
\Phi(M_{VL}) = \int_{-\infty}^{M_{VL}} \phi(M) dM  
\ee
that gives the fraction of galaxies for unit volume, 
present in the sample. Hereafter we adopt the following normalization
for the luminosity function
\be
\label{e411}
\Phi(\infty) = \int_{-\infty}^{M_{min}} \phi(M) dM  = 1 
\ee
where $M_{min} \approx -10 \div -12$ is the fainter galaxy present in 
the available samples. The luminosity factor of Eq.\ref{e410} is useful to
normalize the space density in different VL samples which 
have different $M_{VL}$.


\section{Three dimensional properties}

We have   computed the conditional average density for the  VL samples
with only one cut in absolute magnitude, and we show in Fig.\ref{apmfig2}
the results.
\begin{figure}
\vspace{10cm}
\caption{\label{apmfig2}
The redshift space conditional average density computed for some VL sample of
the SARS redshift survey.
The fractal dimension is $D = 2.1 \pm 0.1$, depending on the 
VL sample used. We have evaluate the errors corresponding to the 
density measurements through the technique of bootstrap resampling (see text)
}
\eef
The fractal dimension is $D =2.1 \pm 0.1$ 
up to
$R_{eff} \sim 40  \hmp$.
In Fig.\ref{apmfig2}  we may recognize three different regimes:
the first one at small distances $r \ltapprox 2 \div 6 \hmp$ shows 
a fluctuating nature and in general a more steeper decay.
This is due to the finite size effect discussed previously:
at these scales we are in the limit $ r \ltapprox \langle \ell \rangle$
as it results from Tab.\ref{tableapm1}. Then, at large scale, $\Gamma^*(r)$
shows a well defined power law decay in the range 
$2 \div 6  \hmp \ltapprox r \ltapprox 30 \div 40 \hmp$ in the 
different samples. The last few points are a noisy because 
the number of points $N(r)$ in the average of eq.\ref{e327} is 
rather small.

We have evaluate the statistical errors corresponding to the 
density measurements.
The standard method for computing the correlation function errors
is through the technique of bootstrap resampling (Ling \etal, 1986). 
We have generated a series of $N = 100 $ bootstrap data sets
of the same size of the original data set by randomly  choosing
    the bootstrap galaxies from 
 the original sample. Each
bootstrap  sample contains 100 randomly selected galaxies.

In general for $ r \ltapprox \langle \ell \rangle$ 
the signal is rather noisy and the bootstrap errors
are large. We have eliminated those points, at small scale,
for which the statistical error is larger than the signal itself.
Moreover we have considered the determination 
of the conditional density only if the average 
is performed over more than 20 points. This is because,
especially at large scales, there are very few points 
which contribute to the average. Moreover the
statistical errors have been computed  
from those   points, which are in the center of
the sample: such a situation may introduce a systematic
effect which may perturb the behaviour of the 
conditional density at large scale.

In order to
 check whether the luminosity incompleteness of the sample for
apparent magnitude brighter than $14.5$  affects substantially
the trends found in Fig.\ref{apmfig2}, we have computed the
conditional average density for the VL samples, with two cuts in distance and 
two in absolute magnitude. These samples  are defined by a lower and an upper
 cut in distance
\begin{figure}
\vspace{10cm}
\caption{\label{apmfig3}
The redshift space 
conditional average density computed for some VL sample of
the SARS redshift survey, with two cuts
in distance and absolute magnitude. 
Such a procedure avoids the luminosity incompleteness
of galaxies with apparent magnitude brighter than $14.5$.
 The fractal dimension is $D = 2.1 \pm 0.1$.
 We have evaluate the statistical errors corresponding to the 
density measurements through the technique of bootstrap resampling (see text)
}
\eef
Even in this case (Fig.\ref{apmfig3}) the fractal dimension 
turns out to be 
$D = 2.2 \pm 0.2$ up to $R_{eff} \sim 20 \hmp$.
This agreement is essentially due to the 
fact that the  galaxies with apparent magnitude 
brighter than $14.5$ represent a small fraction of the 
total sample, and do not perturb the final result.
Clearly in this case the effective volume is smaller and
so  $R_{eff}$.

 As long as the
 space and the luminosity density can be considered independent,
 the normalization of $\Gamma(r)$   in different VL
 samples can be simply done by dividing its amplitude  for the
 corresponding luminosity factor. 
Of course such a normalization is parametric, because it depends on
 the two parameters of the luminosity function $\delta$ and $M^*$
For a reasonable choice of these two
 parameters we find that in different VL samples the amplitude of the conditional 
 density matches quite well  (see Fig.\ref{apmfig4})
 For the  SARS catalog the parameters of the
 luminosity function are $\delta= -1.1$ and $M^* = -19.5$ (Loveday \etal, 1996).  

\begin{figure}
\vspace{10cm}
\caption{\label{apmfig4} 
The redshift space 
spatial conditional average density $\Gamma^*(r)$ 
computed in some 
VL samples of the APM-Stromlo 
and normalized to the 
luminosity factor. } 
\eef


\section{Discussion and conclusions}

In order to understand the reason of the disagreement of
our results with those of Loveday \etal (1996), it is
useful to consider the behavior of the standard 
correlation function for a fractal distribution.
 Following Pietronero (1987), the expression of the 
$\:\xi(r)$ (eq.\ref{e331}) in the case of
fractal distribution, is
\be
\label{x3}
\xi(r) = ((3-\gamma)/3)(r/R_{s})^{-\gamma} -1
\ee
where $\:R_{s}$ (the effective sample 
radius) is the radius of 
the spherical volume where one computes the
average density.
From Eq.\ref{x3}  it follows that

i.) the so-called correlation
length $\:r_{0}$ (defined as $\:\xi(r_{0}) = 1$)
is a linear function of the sample size $\:R_{s}$
\be
\label{x4}
r_{0} = ((3-\gamma)/6)^{\frac{1}{\gamma}}R_{s}
\ee
and hence it is a quantity  
simply related to the sample size. Eq.\ref{x4} explains 
the result of Loveday \etal  (1996) for $r_0$. 
A minor discrepancy is due to the using of weighting schemes
and the treatment of boundary conditions.

ii.) $\:\xi(r)$ is a power law only for 
\be
\label{x6}
((3-\gamma)/3)(r/R_{s})^{-\gamma}  \gg 1
\ee
hence for $\: r \ltapprox r_{0}$: for larger distances
there is a clear deviation from a
power law behavior due to the definition of $\:\xi(r)$.
This deviation, however, is just due to the size of
 the observational sample and does not correspond to any real change
of the correlation properties. It is clear that if one estimates the
 exponent of $\xi(r)$ at distances $r \ltapprox r_0$, one
 systematically obtains a higher value of the correlation exponent
 due to the break of $\xi(r)$ in the log-log plot. 
This is actually the case for the analyses performed so far:
 in fact, usually, 
$\xi(r)$ is fitted with a power law in the 
range $ 0.5 r_{0} \ltapprox  
r \ltapprox 2 r_{0}$.
In this case  one obtains  a systematically higher value of 
 the correlation  exponent. In particular,  
the usual estimation of this exponent
by the  $\xi(r)$ function leads to is $\gamma \approx 1.7$, different 
from $\gamma \approx 1$ (corresponding
to $D \approx 2$) that we found by means of  
the $\Gamma(r)$ analysis (see Sylos Labini \etal, 1997 for
more details).

Our conclusion is therefore that  
the usual methods of analysis are intrinsically inconsistent with
respect to the properties of this galaxy  sample.
The correct statistical analysis of the 
experimental data, performed with the methods of modern Statistical 
Physics, shows that the distribution of galaxies is fractal up to the 
limit of the SARS sample. 
These methods, which are able to identify  
self-similar and non-analytical properties, 
allow us to test the usual homogeneity 
assumption of luminous matter distribution. 
The result is that galaxy distribution in this sample is fractal
with $D = 2.1 \pm 0.1$, and this is in agreement with 
the analyses of different samples (see e.g. Montuori \etal, 1997).
These results have a number of theoretical implications 
which are discussed in more detail in 
  Sylos Labini \etal 1997.

\section*{Acknowledgements}
We warmly thank L. Pietronero for useful discussions and
fruitful suggestions. F.S.L.  is especially grateful
R. Durrer and J-P. Eckmann for various interesting and 
valuable comments
and for their kind hospitality at the 
  D\'ept.~de Physique Th\'eorique of the 
   Universit\'e de Gen\`eve. This work has been partially
supported
by the Italian Space Agency (ASI).


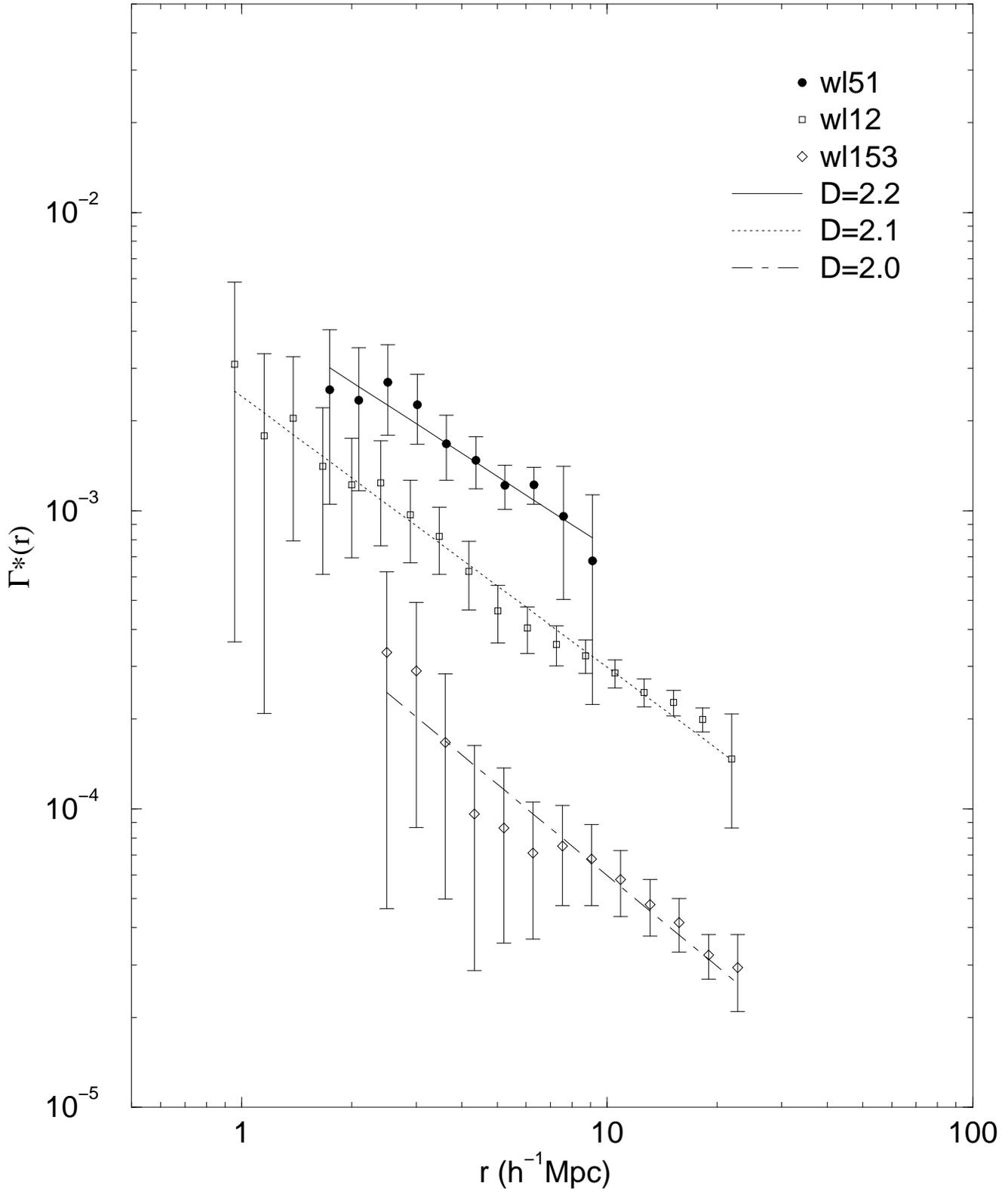
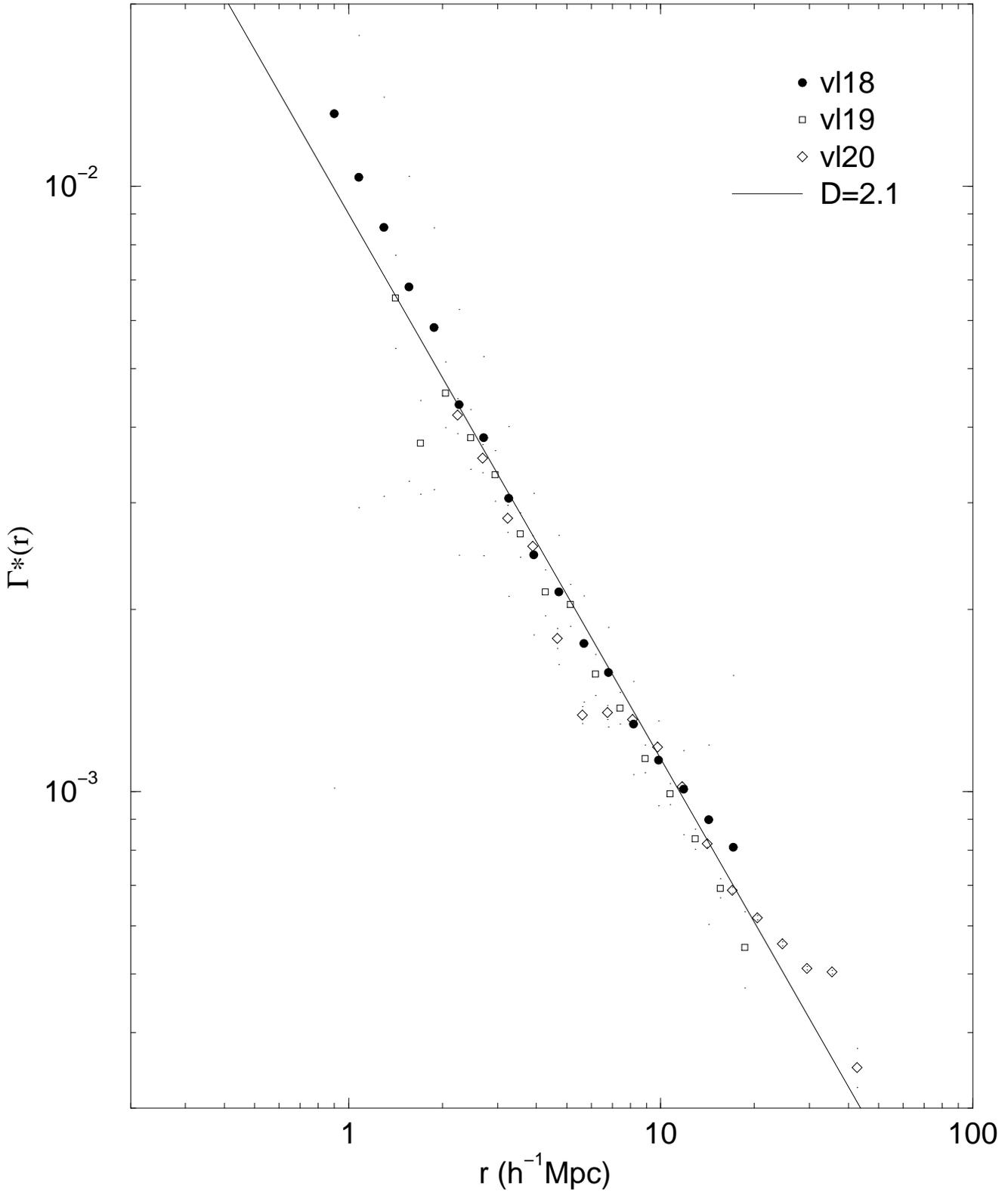
\end{document}